\documentstyle[aps]{revtex}
\begin{document}

\title{The $R_b$ Excess at LEP:\\ Clue to New Physics at the TEVATRON?
\footnote{in the proceedings of the International Symposium on Vector
Boson Self-Interactions, UCLA, February 1995} }

\author{Ernest Ma$^1$ and Daniel Ng$^2$}
\address{$^1$Department of Physics, University of California,\\ Riverside,
California 92521, USA\\
$^2$TRIUMF, 4004 Wesbrook Mall, Vancouver,\\
British Columbia, Canada V6T 2A3}

\maketitle

\begin{abstract}
If the $R_b$ excess at LEP is real, then any explanation in terms of
renormalizable loop corrections leads to important new decay modes of the
$t$ quark and suppresses the $t \rightarrow bW$ branching ratio.  In
the two-Higgs-doublet model, the branching ratio of $Z \rightarrow
b \overline b$ + a light boson which decays itself predominantly into
$b \overline b$ is at least of order $10^{-4}$.

\end{abstract}

\section*{Introduction}

The awesome statistics of the four LEP collaborations have pinned down
with great precision a host of measurable parameters in the standard
model\cite{LEP94}.  The only quantity that shows a possible significant
discrepancy with the theoretical prediction of the standard model is $R_b$
which is defined as
\begin{equation}
R_b \equiv {{\Gamma (Z \rightarrow b \overline b)} \over {\Gamma (Z
\rightarrow {\rm hadrons})}}.
\end{equation}
Using $m_t = 175$ GeV and $m_H = 300$ GeV, the standard model predicts
that $R_b = 0.2158$, whereas LEP obtained $R_b = 0.2202 \pm 0.0020$ if
the similarly defined $R_c$ is assumed to be independent.  If the latter
is fixed at its standard-model value, then $R_b = 0.2192 \pm 0.0018$.
In either case, the excess is about $2\% \pm 1\%$.  If this is taken
seriously, physics beyond the standard model is indicated.

\section*{Two Higgs Doublets}

The simplest extension of the standard model is to have two Higgs doublets
instead of just one.  The relevance of this model to $R_b$ was studied
in detail already a few years ago\cite{hollik91}.  To establish notation,
let the two Higgs doublets be given by
\begin{equation}
\Phi_i = \left( \begin{array} {c} \phi_i^+ \\ \phi_i^0 \end{array} \right)
= \left[ \begin{array} {c} \phi_i^+ \\ 2^{-1/2} (v_i + \eta_i + i\chi_i)
\end{array} \right].
\end{equation}
Let $\tan \beta \equiv v_2/v_1$, then
\begin{eqnarray}
h^+ &=& \phi_1^+ \cos \beta - \phi_2^+ \sin \beta, \\
h_1 &=& \eta_1 \sin \alpha + \eta_2 \cos \alpha, \\
h_2 &=& \eta_1 \cos \alpha - \eta_2 \sin \alpha, \\
A &=& \chi_1 \cos \beta - \chi_2 \sin \beta.
\end{eqnarray}
Note that the $\overline b b A$ and $\overline t b h^+$ couplings involve
the ratio $m_b \tan \beta / M_W$, hence they could be important for large
values of $\tan \beta$.  It was shown\cite{hollik91} that for $\tan \beta
= 70 \simeq 2m_t/m_b$, the $R_b$ excess peaks at about 4\% near $m_A =
m_{h_1} \simeq 40$ GeV for $\alpha = 0$.  However, since $Z \rightarrow
Ah_1$ is not observed, $m_A + m_{h_1} > M_Z$ is a necessary constraint.
We show in Fig.~1 the contours in the $m_{h_1} - m_A$ plane for 3 values
of $R_b$.  It is clear that relatively light scalar bosons are required
if the $R_b$ excess is to be explained.

\begin{flushright}
UCRHEP-T144\\
TRI-PP-95-16\\
April 1995
\end{flushright}

\input psfig

\begin{figure}

\centerline{\psfig{figure=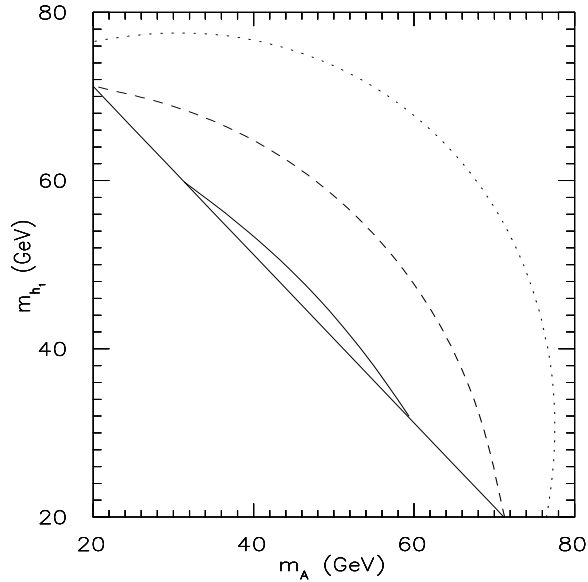,height=3.0in,width=3.0in}}
\vspace {0.2in}

\caption{$R_b=0.2192$ (solid), $0.2174$ (dashed) and $0.2164$ (dotted)
contours in the $m_{h_1} - m_A$ plane for $\alpha=0$ and $\tan \beta = 70$.
The straight line corresponds to $m_A + m_{h_1} = M_Z$.  We have also assumed
$m_{h^\pm} = m_{h_2} = 175$ GeV.}

\end{figure}

For $A(h_1)$ lighter than $M_Z$ and having an enhanced coupling to
$\overline b b$, the decay $Z \rightarrow b \overline b + A(h_1)$
becomes nonnegligible\cite{dzz91}.  As an illustration, we show in
Fig.~2 the branching ratios of these two decays as functions of $m_A$
with the constraint $m_A + m_{h_1} = M_Z + 10$ GeV so that a reasonable
fit to the $R_b$ excess is obtained. It is seen that the sum of these
two branching ratios is at least of order $10^{-4}$.  Once produced,
$A$ or $h_1$ decays predominantly into $b \overline b$ as well.  Hence
this scenario for explaining $R_b$ can be tested at LEP if the
sensitivity for identifying a $b \overline b$ pair as coming from $A$
or $h_1$ in $b \bar{b} b \bar{b}$ final states can be pushed down
below $10^{-4}$.

\begin{figure}

\centerline{\psfig{figure=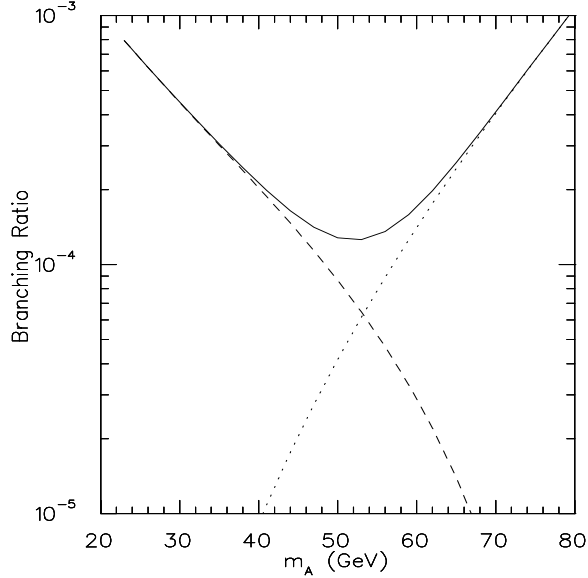,height=3.0in,width=3.0in}}
\vspace{0.2in}

\caption{The branching ratios, $Br(Z \rightarrow b \bar{b} A)$ (dashed)
and $Br(Z \rightarrow b \bar{b} h_1)$ (dotted) and their sum (solid), as
functions of $m_A$ where we take $m_A = m_{h_1} = M_Z + 10$ GeV,
$\tan \beta = 70$, $\alpha = 0$, and $m_{h^\pm} = m_{h_2} = 175$ GeV.}

\end{figure}

\section*{Minimal Supersymmetric Standard Model}

In the Minimal Supersymmetric Standard Model (MSSM), there are two Higgs
doublets, but their parameters are further constrained, hence the
allowed region in the $m_{h_1} - m_A$ plane which gives a large enough
$R_b$ is further reduced by the experimental nonobservation of MSSM
signals at LEP\cite{sopczak94}.

There is of course another possible contribution to $R_b$ in the MSSM:
the $Z \rightarrow b_L b_L$ vertex may be enhanced by the supersymmetric
coupling of $b_L$ to a scalar top quark and a chargino\cite{misc}.
In this case, both of the new particles must again be light, but now $Z$
cannot decay into just one of these particles because of the assumed
conservation of $R$ parity, hence no further constraint is obtainable at LEP.

\section*{Necessary Top Decays}

Since $b_L$ is involved in any enhanced coupling to light
particles in explaining the $R_b$ excess, its doublet partner $t_L$
must necessarily have the same enhanced coupling to related particles.
In the two-Higgs-doublet case, we must have an enhanced $\bar{t} b h^+$
coupling.  Therefore, unless $m_{h^+} > m_t - m_b$, the branching ratio of
$t \rightarrow b h^+$ will dominate over all others.  In particular, the
standard $t \rightarrow b W$ branching ratio will be seriously degraded.
We show in Fig.~3 the branching ratio $Br(t \rightarrow bW)$ as a function
of $m_{h^+}$.  Large values of $m_{h^+}$ are disfavored in this scenario
because the splitting with $A$ and $h_1$ would result in a large
contribution to the $\rho$ parameter\cite{grant95}.
This poses a problem for top production at the TEVATRON because the
number of observed top events is consistent with the assumption that
top decays into $b W$ 100\% of the time.  If that is not so, then top
production must be enhanced by a large factor beyond that of the standard
model.  The two-Higgs-doublet model itself certainly does not provide for
any such mechanism.

\begin{figure}

\centerline{\psfig{figure=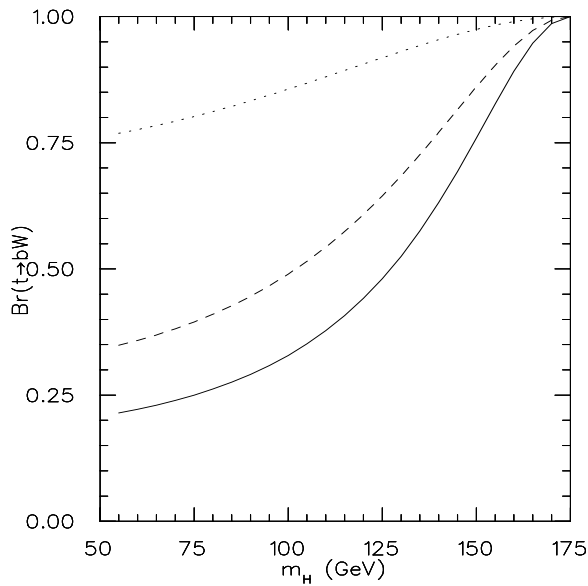,height=3.0in,width=3.0in}}
\vspace{0.2in}

\caption{The branching ratio $Br(t \rightarrow b W)$ as a function of
$m_h^+$ for $\tan \beta = 70$ (solid), 50 (dashed), and 20 (dotted).}

\end{figure}

In the MSSM, if the $R_b$ excess is attributed to a light scalar top quark
and a light chargino, then we should look at the latter's doublet partner
which is in general a linear combination of neutralino mass eigenstates.
At least one of these, {\it i.e.} the Lightest Supersymmetric Particle (LSP),
will be light enough to allow the top quark to decay into it and the
scalar top.  The $\rho$ parameter also serves to disfavor large neutralino
masses in this scenario.  Hence the $t \rightarrow b W$ branching ratio is
again seriously degraded.  Turning the argument around, this means that for
every observed top event, there must be several others which correspond to
the production of supersymmetric particles.  If the $R_b$ excess is really
due to supersymmetry, top decay is the place to discover it!

\section*{Conclusion}

If the $R_b$ excess at LEP is real and we want to explain it in terms of
renormalizable loop corrections, then light particles are unavoidable.
However, these light particles may be produced also in $Z$ decay such as
in the two-Higgs-doublet case, where $Z \rightarrow b \bar{b} +
A~{\rm or}~h_1$ is at least of order $10^{-4}$ in branching ratio.
More importantly, there is necessarily a corresponding top decay into
one of these light particles (such as the scalar top in the MSSM) and
the other particle's doublet partner (the neutralino), which seriously
degrades the $t \rightarrow b W$ branching ratio.  Unless there is
accompanying new physics which enhances the top production by a large
factor at the TEVATRON, this generic explanation of the $R_b$ excess
in terms of light particles does not appear to be viable.

\section*{Acknowledgement}

The work of E.M. was supported in part by the U.S. Department of Energy
under Contract No. DE-AT03-87ER40327.  The work of D.N. was supported by the
Natural Sciences and Engineering Research Council of Canada.

\end{document}